\documentclass[%
 reprint,
 amsmath,amssymb,
 aps,
]{revtex4-2}

\usepackage{graphicx}
\usepackage{dcolumn}
\usepackage{bm}
\usepackage[caption=false]{subfig}

\begin{document}


\title{\textit{Ab initio} theory of magnetism in two-dimensional $1T$-TaS$_2$} 
\author{Diego Pasquier}
\email{diego.pasquier@epfl.ch}
\affiliation{Institute of Physics, Ecole Polytechnique F\'{e}d\'{e}rale de Lausanne (EPFL), CH-1015 Lausanne, Switzerland}
\author{Oleg V. Yazyev}%
\email{oleg.yazyev@epfl.ch}
\affiliation{Institute of Physics, Ecole Polytechnique F\'{e}d\'{e}rale de Lausanne (EPFL), CH-1015 Lausanne, Switzerland}

\begin{abstract}
We investigate, using a first-principles density-functional methodology, the nature of magnetism in monolayer $1T$-phase of tantalum disulfide ($1T$-TaS$_2$).
Magnetism in the insulating phase of TaS$_2$ is a longstanding puzzle and has led to a variety of theoretical proposals including notably the realization of a two-dimensional quantum-spin-liquid phase.
	By means of non-collinear spin calculations, we derive \textit{ab initio} spin Hamiltonians including two-spin bilinear Heisenberg exchange, as well as biquadratic and four-spin ring-exchange couplings.
We find that both quadratic and quartic interactions are consistently ferromagnetic, for all the functionals considered.
Relativistic calculations predict substantial magnetocrystalline anisotropy.
Altogether, our results suggest that this material may realize an easy-plane XXZ quantum ferromagnet with large anisotropy.
\end{abstract}


%
\keywords{Two-dimensional materials, strongly correlated systems, magnetism ,density functional theory}
\maketitle

\paragraph*{Introduction.}
Materials with flat bands near the Fermi level are an ideal platform to study correlated-electrons phenomena. 
There has recently been a renewal of interest in flat-band materials because of the discovery of correlated-insulating phases and superconductivty in twisted bilayer graphene near the so-called magic angle \cite{cao2018unconventional, cao2018correlated}. 
Full characterization and understanding of magnetism in such systems remains a theoretical challenge. 
In certain transition metal dichalcogenides (TMDs), flat bands emerge near the Fermi level due to the occurence of a commensurate charge-density-wave (CDW) phase with star-of-David (SOD) reconstruction \cite{sipos_mott_2008, darancet_three-dimensional_2014, pasquier2018charge, chen2020strong}. 
Among those materials, the most studied one is undoubtedly octahedral tantalum disulfide, which will be the focus of the present work. 

\begin{figure}
  \includegraphics[width=0.8\columnwidth]{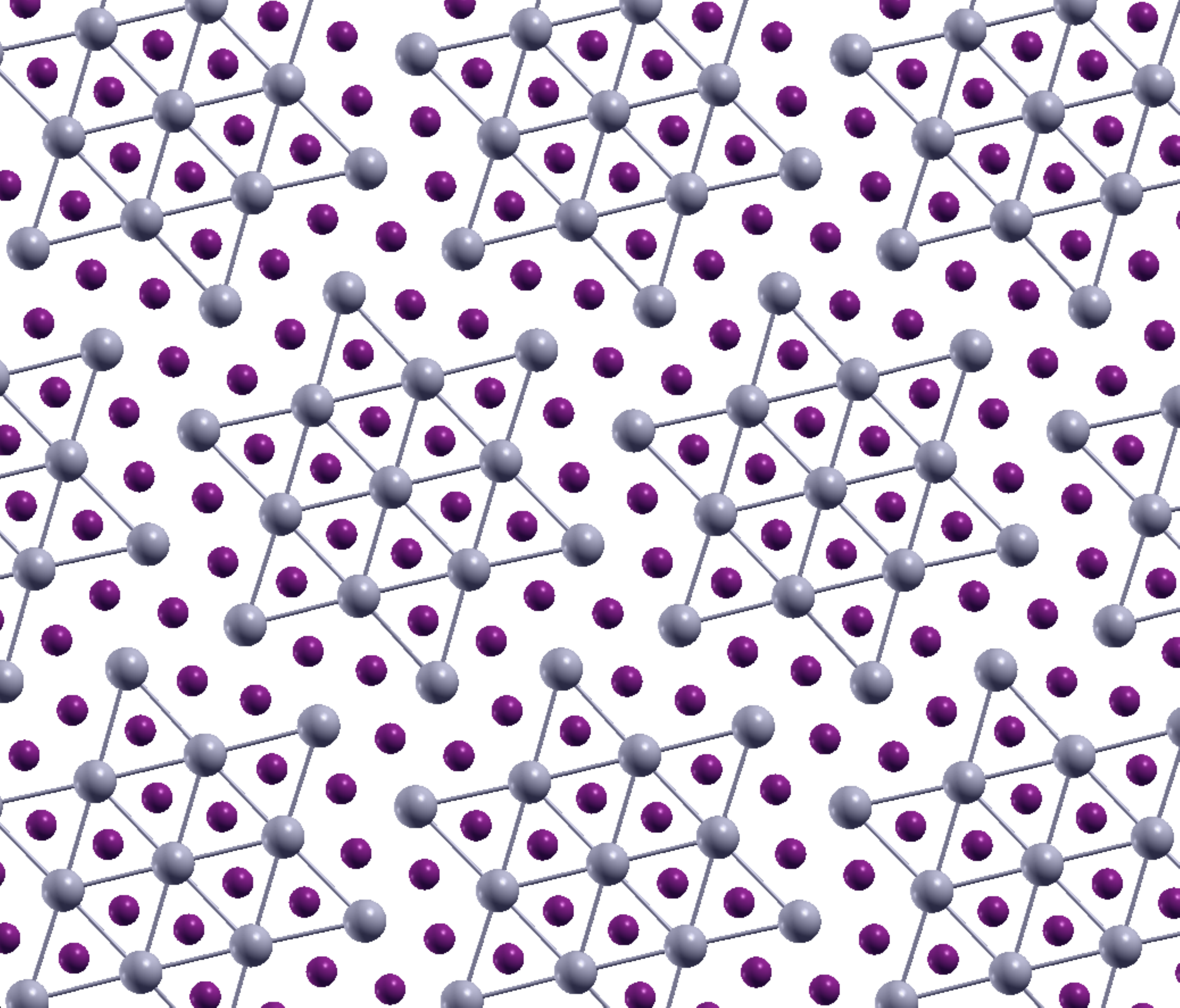}
  \caption{Ball-and-stick representation of TaS$_2$ in the commensurate CDW phase. Grey balls represent Ta atoms, while purple balls stand for S atoms. Only the shortest Ta-Ta bonds were drawn in order to facilitate visualization of the star-of-David pattern.}
  \label{fig:sod}
\end{figure}

The polymorph of tantalum disulfide with distorted octahedral coordination of metal atoms ($1T$-TaS$_2$) is a special member of the family of transition metal dichalcogenides family of materials.
Like many other metallic TMDs \cite{wilson_charge-density_1975, castro_neto_charge_2001, rossnagel2011origin, manzeli20172d, pasquier2019unified}, it undergoes charge-density-wave (CDW) instabilities as the temperature is lowered.
At low temperature, it crystallizes in the so-called SOD phase (illustrated in Fig.\ref{fig:sod}), with $\sqrt{13}\times\sqrt{13}$ periodicity and 13 Ta atoms per supercell.
In this SOD phase, insulating behaviour is observed, which has been commonly attributed to electron localization due to strong correlations \cite{fazekas1980charge, sipos_mott_2008}.
The reason is the following: in the SOD phase, the number of electrons per unit cell is odd, since each Ta atom contributes one conduction $t_{2g}$ electron \cite{pasquier2019crystal}, so conventional band theory would predict a metal.
This argument is clear for the monolayer, but there is actually a caveat for the multi-layer and bulk cases: the stacking is not exactly known, so the three-dimensional character could become important and the exact role of interlayer interactions is unclear \cite{law20171t}.

A longstanding question regarding TaS$_2$ concerns magnetism or, to be more precise, the apparent absence of it \cite{law20171t}, as even the signature of local moment formation is not observed.
Several scenarios have been proposed in the past decades.
The peculiar properties of TaS$_2$ were part of the inspiring data behind Anderson's theory of resonating valence bonds \cite{anderson1973resonating}.
Fazekas and Tosatti suggested that the strong spin-orbit coupling (SOC) of Ta atoms could suppress the magnetic moments \cite{fazekas1979electrical}.
A fluctuating Néel order was proposed in Ref.~\cite{perfetti2005unexpected} and recently, the idea of a spin liquid in TaS$_2$ has gained momentum after Law and Lee proposed that TaS$_2$ might be a rare realization of a quasi-2D quantum-spin-liquid (QSL) phase \cite{law20171t}.
In follow-up work \cite{he2018spinon}, the theory was worked out, yielding the interesting proposal of the possible realization of a spinon Fermi surface in TaS$_2$.
This interesting proposal has triggered experimental effort observe these unconventional states, not only in TaS$_2$ but also in similar materials such as TaSe$_2$ \cite{manas2021quantum, ruan2020imaging}.
A key ingredient in the theory is the presence, in the effective spin Hamiltonian, of a sizeable antiferromagnetic four-spin ring-exchange term beyond the conventional Heisenberg model.

The purpose of the present work is to propose a realistic picture of the nature of magnetism in 2D $1T$-TaS$_2$ derived from first principles, using a density-functional-theory (DFT) framework.
Using non-collinear density-functional calculations, we aim to provide estimates for the magnetic interaction parameters.
In particular, we address whether magnetism in this material is expected to deviate strongly from Heisenberg behavior indeed, as discussed above.
While estimating Heisenberg exchange couplings using DFT is a relatively simple (although delicate) and well-established procedure, the calculation of quartic terms is less common and more challenging.
The approach that we shall adopt here is that of non-collinear spin calculations (see e.g. Refs.~\cite{fedorova2015biquadratic, fedorova2018four}).
The method consists in calculating the total energy for a series of angles between the spins on different sites of the superlattice.
While two-spin bilinear terms lead to a linear dependency on the angle cosine only, quartic terms lead to a quadratic dependency, allowing one to extract the biquadratic and ring-exchange couplings.

\paragraph*{Model Hamiltonian.}
Following He \textit{et al.} \cite{he2018spinon} and neglecting anisotropy for the moment, we write $\mathcal{H}$, the low-energy Hamiltonian describing spin physics in the 2D Mott-insulating phase of $1T$-TaS$_2$:
\begin{equation}\label{eq:hamiltonian}
	\mathcal{H}=\mathcal{H}_\mathrm{Heis}+\mathcal{H}_\mathrm{plaq}+\mathcal{H}_\mathrm{b} \, \, \, ,
\end{equation}
where $\mathcal{H}_\mathrm{Heis}$ is the usual Heisenberg Hamiltonian, describing bilinear spin interactions, and $\mathcal{H}_\mathrm{plaq}$ describes quartic four-spin plaquette terms.
In Eq.~\ref{eq:hamiltonian}, we also explicitly include biquadratic interaction $\mathcal{H}_{b}$.

\begin{figure}
\includegraphics[width=0.8\columnwidth]{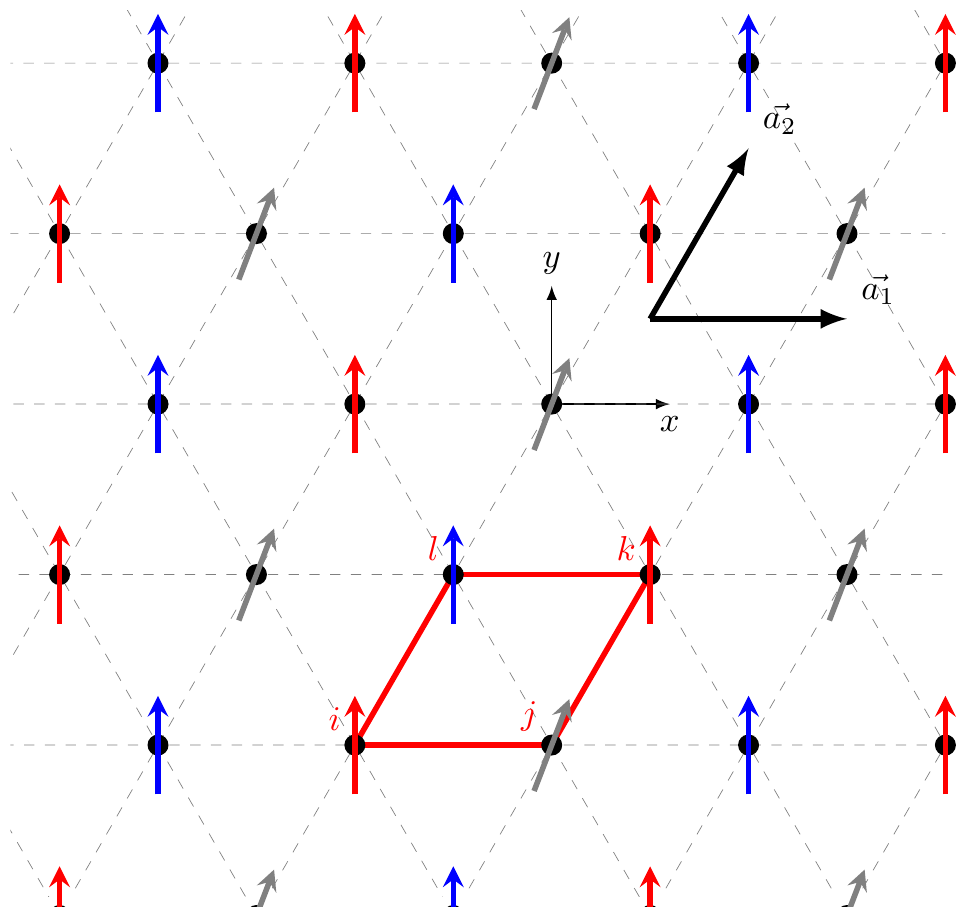}
        \caption{\label{fig:lattice} Schematic representation of the spin lattice employed for the spin-polarized calculations. Each black dot represents a David star with 39 atoms. The $\sqrt{3}\sqrt{13}\times\sqrt{3}\sqrt{13}$ supercell contains three spin sublattices (one of each color). The red diamond illustrates an example of plaquette on the triangular lattice.}
\end{figure}

The Heisenberg Hamiltonian is given by
\begin{equation}\label{eq:heis}
	\mathcal{H}_\mathrm{Heis}=J \sum_{<ij>} \vec{S_i}\cdot\vec{S_j}\, \, ,
\end{equation}
where $<ij>$ denotes nearest-neighbor (NN) sites of the triangular lattice, $J$ is the nearest-neighbor Heisenberg coupling, and $\vec{S_i}$ stands for the $S=1/2$ spin operator at site $i$.
Next-to-nearest-neighbor interactions can be safely neglected, because they are much smaller compared to NN ones.
The biquadratic Hamiltonian is given by $\mathcal{H}_{b}=B \sum_{<ij>} (\vec{S_i}\cdot\vec{S_j})^2$, and the four-spin Hamiltonian $\mathcal{H}_\mathrm{plaq}$ describes higher-order spin-spin interactions and reads:
\begin{equation}\label{eq:plaq}
	\begin{split}
		\mathcal{H}_\mathrm{plaq} & = K \sum_{<ijkl>} (\vec{S_i}\cdot\vec{S_j})(\vec{S_k}\cdot\vec{S_l}) + (\vec{S_i}\cdot\vec{S_l})(\vec{S_j}\cdot\vec{S_k}) \\
		& -(\vec{S_i} \cdot \vec{S_k})(\vec{S_j} \cdot \vec{S_l}) \, \, ,
	\end{split}
\end{equation}
where $<ijkl>$ denotes a plaquette (see Fig. \ref{fig:lattice}), and $K$ is the ring-exchange coupling.
\paragraph*{Methodology.}
All the first-principles calculations presented in this paper were carried out using the Quantum ESPRESSO package \cite{giannozzi2009quantum} .
Lattice parameters and atomic positions were obtained by minimizing forces and stress in the $\sqrt{13} \times \sqrt{13}$ supercell, using the generalized-gradient approximation (GGA) \cite{perdew1986accurate, perdew1996generalized}.
A grid of $8\times8$ k-points was used, with a Marzari-Vanderbilt (MV) \cite{marzari1999thermal} smearing of $0.01$ Ry.
Projector-augmented-wave (PAW) \cite{blochl1994projector} pseudo-potentials from  Ref.~\cite{dal2014pseudopotentials}, including explicitly $s$ and $p$ semicore states for Ta atoms, were used to describe interaction between core and valence electrons.
Plane-wave cutoffs were set to $60$ Ry and $300$ Ry for wave functions and charge density, respectively.

Heisenberg interactions were calculated by comparing total energies with different spin configurations in a $\sqrt{3}\sqrt{13} \times \sqrt{3}\sqrt{13}$ supercell ($2\sqrt{13} \times\sqrt{13}$ for the evaluation of the biquadratic term), using both the local-density approximation (LDA) and the GGA, with and without on-site Hubbard $U$ parameter for Ta $5d$ orbitals.
For DFT+$U$ calculations, we have set $U=2.5$ eV.
For spin-polarized calculations, we have used a smaller MV smearing of $0.001$ Ry, together with a grid of $3 \times 3$ \textit{k}-points in the supercell.
These parameters were chosen to ensure convergence of the exchange couplings.

\begin{figure}
  \includegraphics[width=\columnwidth]{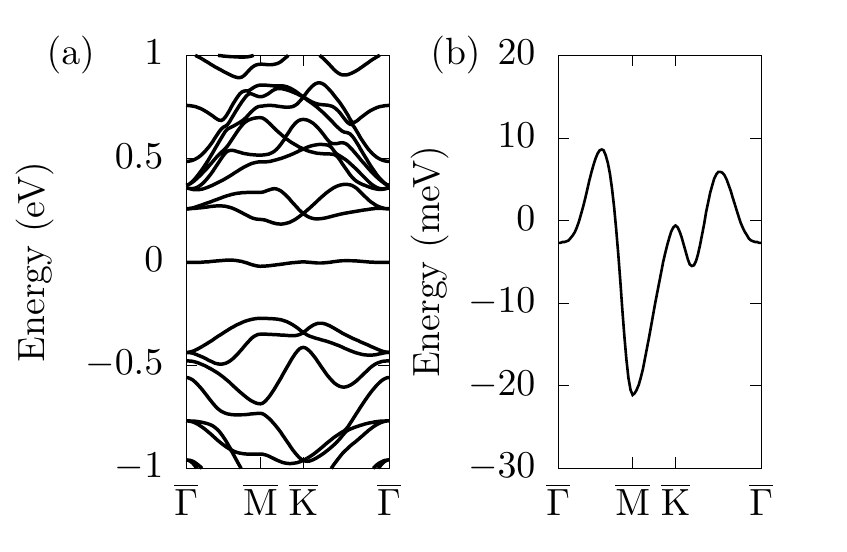}
	\caption{ (a) DFT non-polarized band dispersion of single-layer TaS$_2$ in the star-of-David phase, along the high-symmetry directions of the mini-Brillouin zone. The Fermi energy is set to zero. (b) Dispersion of the half-filled flat band. }
  \label{fig:bands}
\end{figure}

\paragraph*{Structure.}
We begin by optimizing the lattice parameters and atomic positions, in a $\sqrt{13} \times \sqrt{13}$ supercell containing 13 Ta and 26 S atoms.
The resulting structure minimizing forces and stress was found to be the well-known SOD phase, illustrated in Fig.\ref{fig:sod}.
The calculated superlattice parameter was 12.18 \AA\, and the shortest Ta-Ta bonds inside the SOD was 3.18 \AA\ (compared to 3.38 \AA\ in the high-symmetry phase), in line with previous reports \cite{darancet_three-dimensional_2014}.  
Fig.~\ref{fig:bands} shows the noninteracting DFT bands, exhibiting a half-filled narrow band with a bandwidth of $\sim 30$~meV.
In accordance with the existing literature \cite{darancet_three-dimensional_2014, miller2018charge, yi2018coupling}, we also find the corresponding electronic structure insulating when spin polarization is allowed.
Parameters of an effective Hubbard model can be roughly estimated from the DFT results.
Through a Wannier transformation we find a NN hopping of $t\approx 2.5$~meV, and the effective Hubbard interaction can be estimated by examining the correlation gap, which gives $U_\mathrm{eff}\approx 125$~meV with DFT and $U_\mathrm{eff}\approx 330$~meV with DFT+$U$, meaning the $U_\mathrm{eff}/t \sim 132$ ratio sits well in the Mott-Hubbard regime.

\paragraph*{Ab initio parameters: isotropic case.}
\begin{table*}
	\caption{\textit{Ab initio} Heisenberg, biquadratic and ring-exchange parameters (in meV), calculated with different methods and approximations.}\label{tab:heisenberg}
\begin{ruledtabular}
\begin{tabular}{lllllll}
	& $J$ (collinear)   & $J$ & $B$ & $K$ & $J'$ & $K'$ \\
\hline\\
	LDA & $-1.11$ & $-1.15$ & $-0.22$ & $-0.02$ & $-1.12$ & $-0.24$\\
	LDA+$U$ & $-1.19$ & $-1.13$ & $-0.07$ & $-0.18$ & $-1.12$ & $-0.24$\\
	GGA & $-1.62$ & $-1.66$ & $-0.73$ & $-0.04$ & $-1.57$ & $-0.77$\\
	GGA+$U$ & $-1.11$ & $-1.11$ &  $-0.55$ & $-0.03$ & $-1.04$ & $-0.57$\\
\end{tabular}
\end{ruledtabular}
\end{table*}

We turn to the investigation of magnetic properties and examine first the collinear case, neglecting SOC and higher-order interactions.
This will allow us to compare with the values obtained from non-collinear calculations, in order to check the robustness and consistency of the methodology.
We adopt a $\sqrt{3}\sqrt{13} \times \sqrt{3}\sqrt{13}$ supercell (see Fig.~\ref{fig:lattice} ), containing three David stars.
In the calculations, we can force the solution to converge to different spin configurations by appropriately choosing the initial spin orientation in the self-consistent cycle.
The exchange coupling is then given by:
\begin{equation}\label{eq:collin}
        J_{\mathrm{collinear}}= \frac{E_{\uparrow\uparrow\uparrow}-E_{\downarrow\uparrow\uparrow}}{12S^2} \, \, ,
\end{equation}
with $S=1/2$ for TaS$_2$.
In Eq. \ref{eq:collin}, $E_{\uparrow\uparrow\uparrow}$ stands for the energy of the configuration with all the spin parallel to each other, while $E_{\downarrow\uparrow\uparrow}$ is the energy of the configuration in which the spin of one of the three stars points antiparallel to the other two.
Results for different functionals are shown in Table~\ref{tab:heisenberg}.
While the exact values of the parameter are somewhat sensitive to the choice of functional and Hubbard parameter, the qualitative picture remains robust, i.e. coupling is predicted to be ferromagnetic (with a rather small value of $\sim 1$~meV), in line with the cousin materials NbSe$_2$ and NbS$_2$ for which similar values of the parameters were reported \cite{pasquier2018charge, tresca2019charge}. 

In a one-band Hubbard model, the antiferromagnetic exchange would be $J_\mathrm{AFM}=4t^2/U_\mathrm{eff}$, which gives $\approx 0.08~$meV with our estimated DFT+$U$ parameters.
This is an order of magnitude smaller than the estimated $J$, indicating that other mechanisms prevail.
Multi-band Hubbard models can have ferromagnetic kinetic exchange (see Refs.~\cite{penc1996ferromagnetism, huang2020ferromagnetic}) due to third-order (e.g. cyclic) processes, which tend to become dominant in the limit of a very flat band. 
Magic-angle twisted bilayer graphene was recently also proposed to be a ferromagnetic Mott insulator \cite{seo2019ferromagnetic}, driven by a large intersite direct Coulomb exchange that dominates the kinetic $J_\mathrm{AFM}$, the latter being small due to the bands' flatness.

We then turn out attention the non-collinear case.
Let us define $\alpha$ as the angle between the spin direction of different stars in the supercell, as shown in Fig. \ref{fig:lattice}.
We can write the total energy as a function of $\alpha$ in the following way:
\begin{equation}
        E(\alpha) = E_0 + E_1\mathrm{cos}(\alpha) + E_{2}\mathrm{cos}^2\alpha \, \, .
\end{equation}
The two-spin and four-spin exchange can then be extracted from the $E_1$ and $E_2$ parameters as:
\begin{equation}\label{eq:params}
        J= \frac{2E_1 -E_2+6BS^4}{12S^2} \, \, , K= \frac{E_2-6BS^4}{6S^4} \, \,.
\end{equation}
The system of equations \ref{eq:params} is underdetermined, so we need to first determine $B$ using a different supercell.
This can be done by considering the $2\sqrt{13}\times\sqrt{13}$ supercell for which the quadratic term $E_2^\mathrm{2\times1}$ depends only on $B$, with $B=E_2^\mathrm{2\times1}/4S^4$.

\begin{figure}
  \includegraphics[width=\linewidth]{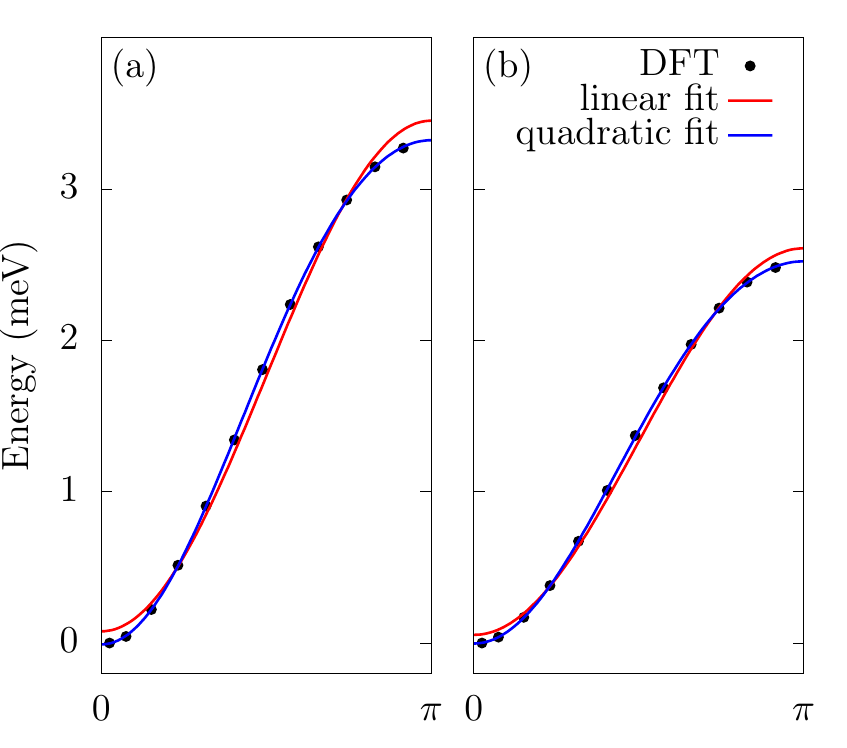}
 
	\caption{Total energy as a function of the angle between spins, calculated with GGA+$U$, in the (a) $\sqrt{3}\sqrt{13}\times\sqrt{3}\sqrt{13}$ and (b) $2\sqrt{13}\times\sqrt{13}$ supercells. Fits wit linear and quadratic functions are also presented.}
  \label{fig:fit}
\end{figure}

Fig.~\ref{fig:fit} shows the calculated total energies in the GGA+$U$ case for a set of selected angles in both $\sqrt{3}\sqrt{13}\times\sqrt{3}\sqrt{13}$ and $2\sqrt{13}\times \sqrt{13}$ supercells.
Fig.~\ref{fig:fit} also shows the least-square fits with linear and quadratic functions, showing that the quadratic form allows one to almost perfectly fit the DFT results, whilst the linear fit, which corresponds to a mapping to a simple Heisenberg model, shows deviations.
However, it is evident from Fig.~\ref{fig:fit} that the deviations from the Heisenberg model are only modest.
This is reflected by the ratio $E_2/E_1 \sim 0.1$ that shows that contributions beyond bilinear two-spin terms are subdominant. 
Using LDA, the ratio found is even smaller ($E_2/E_1\sim 0.05$). 
Although GGA is in general an improvement compared to LDA, the formulation of non-collinear magnetism with gradient corrections is ambiguous, so we would expect the LDA results to be more rigorous.
We have also checked that higher-order terms (i.e. higher power in the expansion) are negligible.
This is one of the central results of the present work, that first-principles calculations predict no substantial deviations from Heisenberg physics, i.e. from a Hamiltonian truncated to quadratic terms in the spin operators.
It was however not obvious \textit{a priori} since it had been hypothesized otherwise in theoretical work, and was therefore worth checking.
Our finding can be understood in light of the large $U_\mathrm{eff}/t \gg 1$ ratio found, which is different than what was often taken in model Hamiltonians \cite{perfetti2005unexpected, perfetti2008femtosecond, he2018spinon}.

In Table~\ref{tab:heisenberg}, values for $J$ $B$, and $K$ obtained with different functionals are reported.
We also show as a sensitivity analysis $J'$ and $K'$, which were obtained by mapping the DFT results to a model without biquadratic exchange.
Bilinear Heisenberg terms obtained from collinear and non-collinear calculations are consistent, demonstrating the soundness of our methodology.
Biquadratic and ring exchange are also found to be ferromagnetic for both LDA and GGA, with or without the Hubbard correction $U$.
Table~\ref{tab:heisenberg} shows the extracted value of $K$ is very sensitive to the choice of model (i.e. with or without biquadratic coupling) and also quite sensitive to the choice of functional..
Nevertheless, the sign of the coupling is robust and indicates a QSL phase is not compatible with the DFT results because unlike other combinations \cite{grover2010weak}, the scenario with both negative $J$ and $K$ is not predicted to host such state.

\paragraph*{Spin anisotropy.}
Our results suggest that a ferromagnetic state might be observed at low temperature in monolayer TaS$_2$.
In two-dimensional systems, a finite Curie temperature requires spin anisotropy, driven by the SOC.
We calculated, using the LDA(+$U$)+SOC functional, the in-plane and out-of-plane bilinear couplings denoted $J_{x}$ $J_y$ and $J_z$, with $z$ the out-of-plane direction.
We consider here the quadratic Hamiltonian $\mathcal{H}=\sum_{<ij>}(J_xS_i^xS_j^x+J_yS_i^yS_j^y+J_zS_i^zS_j^z)$, neglecting possible anisotropy of the quartic terms.
The methodology used is the same as for the evaluation of isotropic $J$ using collinear calculations, except that we consider three different spin orientations.
For these non-collinear calculations including the SOC, we have not used the GGA because of possible convergence issues in that case.

Extracted parameters are shown in Table~\ref{tab:gamma}.
In-plane couplings $J_x$ and $J_y$ are equal, indicating in-plane isotropy.
We observe smaller couplings in the relativistic case compared to isotropic $J$, that we understand as due to the reduced magnetic moment induced by the SOC ($\approx$0.82$\mu_\mathrm{B}$/SOD with LDA+$U$).
The total energy difference between the ferromagnetic state with $x$ and $y$ polarization is negligibly small, $\mathcal{E}_x^{\mathrm{FM}}-\mathcal{E}_y^{\mathrm{FM}} < 0.1$~$\mathrm{\mu}$eV/SOD, which we interpret as evidence for almost if not perfect isotropy.
On the other hand, the out-of-plane coupling $J_z$ is significantly smaller, giving a negative anisotropy term $\gamma=J_z/J_x-1=-0.3$ in DFT+$U$. 
The exchange anistropy $\gamma$ was found to be insufficient to account for the calculated magnetocrystalline anistropy $\mathcal{E}_z^{\mathrm{FM}}-\mathcal{E}_x^{\mathrm{FM}} \approx 0.27$~meV/SOD in LDA+$U$.
The simplest way to account for the remaining difference is through an on-site anisotropy term $D\sum_i(S_i^z)^2$ in the Hamiltonian.
The corresponding extracted values for $D$ are shown in Table~\ref{tab:gamma}.

Our first-principles results thus indicate that, at the DFT(+$U$) level of theory, 2D TaS$_2$ in the SOD phase maps into an easy-plane quantum ferromagnet.
For the XXZ model with easy-plane anisotropy, the Mermin-Wagner theorem applies meaning that no finite-temperature second-order phase transition is expected \cite{mermin1966absence}.
This is unlike the easy-axis case for which a finite Curie temperature is predicted due to the opening of a gap in the spin-wave dispersion (see e.g. Ref.~\cite{lado2017origin} in the case of CrI$_3$).
Classical Monte-Carlo simulations of the easy-axis XXZ model on a two-dimensional lattice show a Berenzinskii-Kosterlitz-Thouless transition with $T_\mathrm{BKT}\sim JS^2$ \cite{cuccoli1995two}.
Results for the quantum model were less conclusive, since large fluctuation in the extreme quantum limit ($S=1/2$) may reduce the effective anisotropy and exchange coupling magnitude \cite{cuccoli1995quantum}.
We have found that the four-spin terms are also predicted to be ferromagnetic (and rather small), and in that case the expected effect of the ring-exchange term is to merely renormalize the spin wave dispersion \cite{owerre2013spin}.

\begin{table}
	\caption{\textit{Ab initio} relativistic Heisenberg and on-site anisotropy parameters (in meV), and corresponding exchange anisotropy $\gamma$. }\label{tab:gamma}
\begin{ruledtabular}
\begin{tabular}{llllll}
	& $J_{x}$ & $J_y$ & $J_z$ & $D$ & $\gamma$ \\
\hline\\
	LDA+SOC & $-0.94$ & $-0.94$ & $-0.75$ & $0.19$ & $-0.20$ \\
	LDA+$U$+SOC & $-0.78$ & $-0.78$ &$-0.54$ & $0.37$ & $-0.30$\\
\end{tabular}
\end{ruledtabular}
\end{table}

\paragraph*{Discussion.}
Magnetism in the SOD phase of TaS$_2$ is a longstanding question, which as led to several interesting proposals.
Theoretical understanding has been so far hindered by two factors.
First, experimental work probing magnetism was only done on layered form of TaS$_2$.
Uncertainty on the exact stacking remains and the approximation of a quasi-2D material is questionable.
Secondly, an \textit{ab initio} model Hamiltonian for spins was not known, and parameters were mostly derived from simplified one-band Hubbard models.
Our results differ significantly from these models, as we have found opposite sign for both two-spin and four-spin couplings (i.e we have found both terms to be ferromagnetic).
In theory, our results are derived from first principles so they should be more robust compared to a Hubbard model studied perturbatively to derive a spin Hamiltonian.
However, in the presence of strong correlations, we cannot exlude that the approximations of the DFT+$U$ method break down and that it fails qualitatively, even though it usually performs well for magnetism in various correlated systems.

The present work suggests that single-layer TaS$_2$ could realize a 2D easy-plane ferromagnet.
This could help interpret future experiments, not only on TaS$_2$ but also on similar materials such as TaSe$_2$, although the generalizability of the present results to other materials remains to be established.
If confirmed, that would also have bearing on the understanding of bulk TaS$_2$, as it would mean that a quasi-2D spin-liquid cannot explain the absence of magnetic ordering observed.
There is growing evidence that interlayer interactions are more important than initially thought and that this material might actually be quasi-one-dimensional rather than quasi-2D \cite{darancet_three-dimensional_2014, ngankeu_quasi-one-dimensional_2017, butler2020mottness}.
The origin of the observed insulating behaviour is still unclear and might therefore be explained from a different perspective than a Mott insulator \cite{lee2021distinguishing}.
Recently, it was proposed that bulk TaS$_2$ could actually be a simple band insulator, with gap formation driven by the stacking pattern \cite{ritschel2018stacking, lee2019origin}.

\vskip 0.5cm
\noindent
\bibliographystyle{apsrev4-1}
\bibliography{tas2.bib}
\end{document}